\documentclass[preprint,11pt]{JHEP3}

\JHEPspecialurl{http://jhep.sissa.it/JOURNAL/JHEP3.tar.gz}

\usepackage{epsfig,multicol,amsmath}

\usepackage{epstopdf}
\usepackage{bm}  
\usepackage{amsmath}     
\usepackage{epsfig,epsf}
\usepackage{graphicx}
\usepackage{rotating}
\usepackage{cite}

\def\beq{\begin{equation}}
\def\eeq{\end{equation}}
\def\bea{\begin{eqnarray}}
\def\eea{\end{eqnarray}}

\def\eq#1{{Eq.~(\ref{#1})}}
\def\fig#1{{Fig.~\ref{#1}}}

\newcommand{\Lb}{\left(}
\newcommand{\Rb}{\right)}
\parskip=\itemsep               

\newcommand{\nn}{\nonumber}

\newcommand{\h}{\frac{1}{2}}

\def\pom{{I\!\!P}}
\def\reg{{I\!\!R}}
\title{Soft interaction model and the LHC data.}
\author{\Large  E. Gotsman$^{a}$\thanks{Email:
gotsman@post.tau.ac.il.}\,, E. Levin$^{a,b}$\thanks{Email:
leving@post.tau.ac.il}\,\,and\,\,U. Maor$^{a}$\thanks{Email: maor@post.tau.ac.il.}\, 
\\
a)\,  \,Department of Particle Physics, School of Physics and Astronomy,
Raymond and Beverly Sackler
 Faculty of Exact Science, Tel Aviv University, Tel Aviv, 69978, Israel\\ 
b)\,\,Departamento de F\'\i sica, Universidad T\'ecnica Federico Santa 
Mar\'\i a, Avda. Espa\~na 1680\\ and Centro 
Cient??fico-Tecnol$\acute{o}$gico de Valpara\'\i so,Casilla 110-V, 
Valparaiso, Chile\\

\\}

\abstract{    Most models for soft interactions which were proposed prior to the 
measurements at the LHC, are only marginally compatible with LHC data, our
GLM model has the same deficiency.  
 In this paper we investigate  possible causes of the problem, 
by considering separate fits to the high energy ($W > 500\, GeV$), and low 
energy ($W < 500\, GeV$) data. Our new results are moderately higher than our 
previous predictions. Our results for total and elastic cross sections
are systematically lower that the recent Totem and Alice published values, 
while our 
results for the inelastic and forward slope agree with the data.
If with additional experimental data,  the errors are reduced, while 
 the central cross section values remain unchanged, we will need to 
reconsider the physics on which our model is built.}

\dedicated{PACS: 13.85.-t, 13.85.Hd, 11.55.-m, 11.55.Bq}
\preprint{TAUP -2145/12\\
{\tt }\\
\today}
\begin{document}
\section{Introduction}
The new LHC data on soft interaction scattering at high energy 
(see Refs.\cite{ALICE,ATLAS,CMS,TOTEM}) is only marginally compatible with 
updated Pomeron models\cite{GLM,KAP,KMR,OST}, which have been successful 
in reproducing the cross section data in the lower energy range. 
The implication is that, as it stands, 
our understanding of long distance physics at very high energies 
is limited. From an optimistic point of view this may imply no more than 
the need to adjust the Pomeron model's parameters. 
From a pessimistic point of view, this 
may suggest the need for a comprehensive 
revision of the main ingredients of  Pomeron models 
applied to high energy soft interactions.

Specifically, the model we propose \cite{GLM}  is built using an input 
Pomeron 
with a relatively large fitted intercept 
$\Delta_{\pom} = \alpha_{\pom} - 1 = 0.2$ 
and exceedingly small slope $\alpha_{\pom}^{\prime} = 0.02\,GeV^{-2}$. 
These values are 
in accord with AdS-CFT correspondence\cite{BST,HIM,COCO,BEPI,LMKS}. 
Note that in N=4 SYM\cite{BST}, 
$\Delta_{\pom} = 1 - 2/\sqrt{\lambda} \approx 0.1-0.3$,
corresponding to experimental estimates based on 
multiparticle production, as well as HERA DIS data\cite{LEPO} 
in which $\lambda \approx 5-9$. The other  basic ingredients of our model 
are the large Good-Walker(GW)\cite{GW} contribution to diffraction 
production, and a small Pomeron self interaction. 
Both are direct consequences of the AdS-CFT correspondence.

If the present central values of the LHC cross section data points, 
which have relatively large errors, do not change significantly  
with 
the forthcoming, better statistics, measurements. 
This  would suggest either of two extreme options: 
\newline 
1) The new LHC data do not 
support the main theoretical concepts of our model. 
This may stem from our reliance on  Reggeon 
calculus and pQCD, which led to a single Pomeron model, rather than 
the traditional distinction between a soft and a pQCD hard Pomeron,
and/or from our realization of AdS-CFT N=4 SYM ideas.
\newline
2)  Our procedure for adjusting parameters may be deficient, requiring
a more sophisticated data analysis which may yield  
satisfactory results. We note that the fitted data base \cite{GLM} 
contains no LHC data. Moreover,  
the low energy ($W < 500 GeV$) total, elastic 
and diffractive cross sections which constitute the major portion of the 
fitted data points have rather small errors.  
Consequently, our fitting procedure is not well balanced as the 
main contribution to our $\chi^2/d.o.f.$ stems from the low energy data.
\newline
An alternate, and probably a more realistic option would be based 
on elements originating from the above two propositions.

In this paper we check the second option. To this end 
we removed the low energy data and only fitted  
 the high energy data  ($W > 500 GeV$), 
including the available LHC soft cross section data points, 
so as to determine the Pomeron parameters. 
Having adjusted these parameters,  
we tuned the value of the Reggeon-proton vertex, which enabled us 
to obtain a smooth cross section behaviour through the ISR-LHC energy 
range. We hope that this exercise will 
clarify to what extent our model has intrinsic deficiencies,  
or do we  just have  a technical problem in the procedure 
for adjusting our free parameters.

The results of this paper can be summarized as follows: 
First, we show that in spite of the fact that the values of the parameters,   
extracted from our current fitting, are  different from our 
previous values, the overall picture remains unchanged. 
Second, our updated total and elastic cross sections
are slightly lower than the published TOTEM values\cite{TOTEM}, but still 
within the relatively large experimental error bars. Should future LHC 
measurements confirm the present TOTEM values, 
we will need to revise our dynamic picture for soft scattering.

\section{Our Model}

The ingredients and formulae of our model have been published (see 
Ref.\cite{GLM}). However, in order to produce a self contained 
presentation, we start with a brief overview of our formalism.

As we have mentioned, one of our main input assumptions is the 
GW mechanism\cite{GW}, 
which plays a significant role in the calculation of the  
eikonal shadowing corrections. To this end we 
took into account a two channel formalism in which 
we introduced two eigen wave functions, $\psi_1$ and $\psi_2$, which 
diagonalize the 2x2 interaction matrix ${\bf T}$,
\beq \label{2CHM}
A_{i,k}=<\psi_i\,\psi_k|\mathbf{T}|\psi_{i'}\,\psi_{k'}>=
A_{i,k}\,\delta_{i,i'}\,\delta_{k,k'}.
\eeq
In this representation the two observed states are an hadron 
(a nucleon in our calculations), 
denoted by the wave function $\psi_h$ 
and a diffractive state $\psi_D$. Note that, we replaced 
the rich population of the diffractive Fock states by 
a single state with unknown mass.
This representation provides a considerable 
simplification of our calculations at the price of not being able 
to calculate the mass dependence of GW diffraction production.  
The two observed states can be written in the form 
\beq \label{2CHM31}
\psi_h=\alpha\,\psi_1+\beta\,\psi_2\,,\,\,\,\,\,\,\,\,\,
\psi_D=-\beta\,\psi_1+\alpha \,\psi_2\,,
\eeq
where, $\alpha^2+\beta^2=1$.
Using \eq{2CHM}, we can rewrite the s-channel unitarity constraints in the form
\beq \label{UNIT}
2\,\mbox{Im}\,A_{i,k}\left(s,b\right)=|A_{i,k}\left(s,b\right)|^2
+G^{in}_{i,k}(s,b),
\eeq
where, $G^{in}_{i,k}$ is the contribution of all 
non GW inelastic processes.

In a general solution of \eq{UNIT} 
\beq \label{2CHM1} A_{i,k}(s,b)=i \Lb 1 
-\exp\Lb - \frac{\Omega_{i,k}(s,b)}{2}\Rb\Rb, \eeq \beq \label{2CHM2} 
G^{in}_{i,k}(s,b)=1-\exp\Lb - \Omega_{i,k}(s,b)\Rb, 
\eeq 
in which $\Omega_{i,k}$ are arbitrary. 
In the eikonal approximation $\Omega_{i,k}$ 
are real  and amplitude $A_{i k}$ are pure imaginary.  In general we 
have 4 $A_{i,k}$ amplitudes, however, for pp and 
$\bar{p}p$ $A_{1,2}\,=\,A_{2,1}$. 
  From \eq{2CHM2} we deduce that the 
probability that the initial state $(i,k)$ remains intact 
during the interaction, is $P^S_{i,k}=\exp \Lb - \Omega_{i,k}(s,b) \Rb$.

The input opacity $\Omega_{i,k}(s,b)$ 
corresponds to an exchange of a single bare Pomeron.  
\beq \label{Born}
\Omega_{i,k}(s,b) \,\,=\,\, g_i(b)\,g_k(b)\,P(s). 
\eeq
$P(s)\,=\,s^\Delta$ and $g_i(b)$ is the 
Pomeron-hadron vertex parameterized in the form: 
\beq \label{GP} g_i\Lb 
b\Rb\,=\,g_i\,S_i(b)\,=\,\frac{g_i}{4 \pi}\,m^3_i \,b\,K_1\Lb m_i b\Rb. 
\eeq 
$S_i(b)$ is the Fourier transform of $\frac{1}{(1 + q^2/m^2_i)^2}$, 
where, $q$ is the transverse momentum carried by the Pomeron. 
In our calculations 
we assume that the slope of the Pomeron trajectory is\footnote{Actually, 
we sum the diagrams for the Pomeron interaction considering $\alpha'_\pom = 0$,
 but introduce $\alpha'_{\pom}$ for the Pomeron exchange. Since, the output
 of our fit gives a small value of $\alpha'_{\pom} \approx 0.02
 \,GeV^{-2}$, we consider that this procedure is 
 justified {\it a posteriori}.}
$\alpha'_\pom = 0$. 
This is compatible with the exceedingly small fitted value 
of $\alpha_{\pom}^{\prime}$, and in accordance with N=4 SYM \cite{BST}.

In our model\cite{GLM}, the Pomeron's Green function that 
includes all enhanced diagrams is approximated using the MPSI 
procedure\cite{MPSI}, in which a multi Pomeron interaction
(taking into account only triple Pomeron vertices) is          
approximated by large Pomeron loops of rapidity size of $\ln s$.

We obtain 
\beq \label{GFPEN}
 G_{\pom}\Lb Y\Rb\,\,=\,\,1 \,-\,\\exp\Lb \frac{1}{T\Lb Y\Rb}\Rb\,
\frac{1}{T\Lb Y\Rb}\,\Gamma\Lb 0,\frac{1}{T\Lb Y\Rb}\Rb, 
\eeq
in which:
\beq \label{ES11}
T\Lb Y \Rb\,\,\,=\,\,\gamma\,e^{\Delta_{\pom} Y}.
\eeq
$\Gamma\Lb 0, 1/T\Rb$ is the incomplete gamma function
(see formulae {\bf 8.35} in Ref.\cite{RY}).

Summing the net diagrams \cite{GLM}, we replace $g_i(b)$ by a more 
complicated 
vertex function which, together with the enhanced diagrams, results in 
the following expression for $\Omega_{i,k}(s,b)$:
\beq \label{FIMF}
\Omega^{i,k}_{\pom}\Lb Y; b\Rb\,\,\,= \,\,\, \int d^2 b'\,
\,\,\,\frac{ g_i\Lb\vec{b}'\Rb\,g_k\Lb\vec{b} - \vec{b}'\Rb
\,\Big( 1/\gamma\, G_{\pom}\Lb T(Y)\Rb\Big)}
{1\,+\,\Lb G_{3\pom}/\gamma\Rb G_{\pom}\Big(T(Y)\Big)\,\left[g_i\Lb
\vec{b}'\Rb + g_k\Lb\vec{b} - \vec{b}'\Rb\right]}.
\eeq
$G_{3\pom}$ is the triple Pomeron vertex, 
and $\gamma^2 = \int \frac{d^2 k_t}{4 \pi^2} G^2_{3 \pom}$. 
 Note  we consider $\gamma$ as an
 independent parameter in our  fit to the data.

For the elastic amplitude we have:
\beq \label{ES}
a_{el}(b)\,=\,\Lb \alpha^4 A_{1,1}\,
+\,2 \alpha^2\,\beta^2\,A_{1,2}\,+\,\beta^4 A_{2,2}\Rb. 
\eeq 
For diffraction production we introduce an additional contribution due 
to the Pomeron enhanced mechanism which is non GW. 
For single diffraction we have (see Fig.1a):
\bea \label{SD}
A^{sd}_{i; k, l}\,\,\,&=&\,\,\int d^2 b'\,2\,\Delta\,\,
\Big(\frac{G_{3\pom}}{\gamma} \,
\frac{1}{\gamma^2}\Big)\,g_i\Lb \vec{b} - 
\vec{b}',m_i\Rb\,\,g_l\Lb  
\vec{b}',m_l\Rb\,g_k\Lb  \vec{b}',m_k\Rb \nn\\
&\times &\,\,Q\Lb g_i,m_i,\vec{b} - 
\vec{b}',Y_m\Rb\,\,Q\Lb g_k,m_k, \vec{b}',Y - 
Y_m\Rb\,\,Q\Lb g_l,m_l, \vec{b}',Y - Y_m\Rb,
\eea
where,
\beq \label{SD1}
Q\Lb g, m, b; Y\Rb \,\,=\,\,\frac{G_{\pom}\Lb Y\Rb}{ 1\,\,
+\,\,\Lb G_{3\pom}/\gamma\Rb\,g\,G_{\pom}\Lb Y\Rb \,S\Lb b, m \Rb}.
\eeq
 The structure of \eq{SD} can be understood from \fig{dif}-a.
 $ Q\Lb g, m, b; Y\Rb $ describes the sum of the  `fan' Pomeron diagrams. 
 As shown in  \fig{dif}-a, we have one cut Pomeron in \fig{dif}-a, 
 which we express through the Pomeron without a cut, using the AGK cutting
 rules \cite{AGK}.

For double diffraction we have (see Fig.1b):
\bea \label{DD}
A^{dd}_{i,k}\,\,&=&\,\,\int d^2 b'\,  4 \,g_i\Lb \vec{b} - \vec{b}',m_i\Rb\,\,g_k\Lb  \vec{b}',m_k\Rb\nn\\
&\times &\, \,Q\Lb g_i,m_i,\vec{b} - \vec{b}',Y - 
Y_1\Rb\,e^{2 \Delta\,\delta Y} \,Q\Lb g_k,m_k, \vec{b}', Y_1 - \delta Y\Rb.
\eea

 This equation is illustrated in \fig{dif}-b,
 which displays all  ingredients of the equation.
We express each of two cut Pomerons through the Pomeron without
a cut, using the AGK cutting rules \cite{AGK}.

\eq{SD} and \eq{DD} are the simplifications of the exact formulae of 
Ref.\cite{GLM}, which correspond to the diagrams of \fig{dif}. We 
checked that they approach the values of the exact formulae reasonably 
well, within $5 - 10\%$.
  
For single diffraction,  
$ Y = \ln\Lb M^2/s_0\Rb$, where, $M$ is the SD mass. 
For double diffraction, 
$ Y - Y_1 = \ln\Lb M^2_1/s_0\Rb$ 
and $ Y_1 - \delta Y = \ln \Lb M^2_2/s_0  \Rb$, 
where $M_1$ and $M_2$ are the masses of two bunches of hadrons 
produced in double diffraction.
$s_0 $ is the minimal produced mass, which is about $1 \,GeV$.
\FIGURE[ht]{
\centerline{\epsfig{file=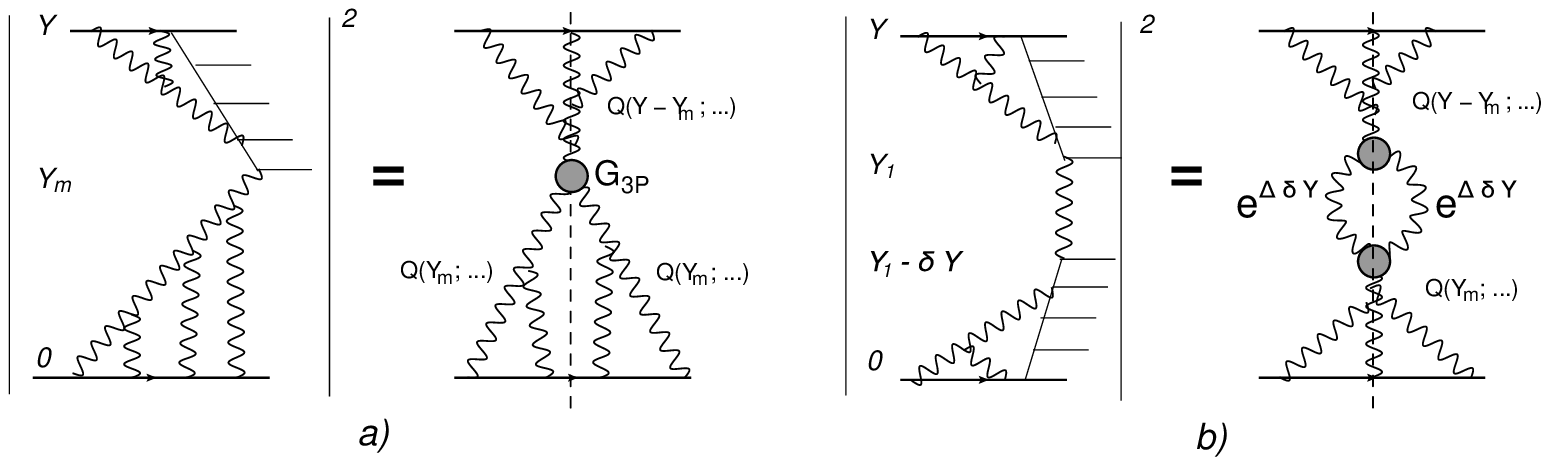,width=140mm,height=40mm}}
\caption{The set of the diagrams for single (\fig{dif}-a) 
and double (\fig{dif}-b) diffraction. 
The wave like lines denote  Pomerons. The solid horizontal lines 
correspond to  interacting hadrons. The vertical dashed line denotes the cut that corresponds to produced particles.}
\label{dif}}

The integrated cross section of the SD channel  
is written as a sum of two terms:
the GW term, which is equal to
\beq \label{FSSDGW}
\sigma^{GW}_{sd}\,\,=\,\,\int\,\,d^2b\,
\Big{|} \alpha\beta\{-\alpha^2\,A_{1,1}
+(\alpha^2-\beta^2)\,A_{1,2}+\beta^2 \,A_{2,2}\}\,\Big{|}^2.
\eeq
$A_{i,k}$ are given by \eq{2CHM1}.
The second term describes diffraction production due to non GW  
mechanism:
\bea\label{FSSDLM}
&&\sigma^{\mbox{nGW}}_{sd}\,\,=\,\,\,\,2 \int d Y_m
\int d^2 b \,\\
&&\left\{\,\alpha^6\,A^{sd}_{1;1,1}\,
e^{- \Omega_{1,1}\Lb Y;b\Rb}\,\,+\,\alpha^2\beta^4 A^{sd}_{1;2,2}\,
e^{- \Omega_{1,2}\Lb Y;b\Rb} + 2\,\alpha^4\,\beta^2 \,A^{sd}_{1;1,2}\,
e^{- \h\Lb \Omega_{1,1}\Lb Y;b\Rb
+ \Omega_{1,2}\Lb Y; b \Rb\Rb} \right. \nn\\
&&\left.\,\,+\,\,\beta^2\,\alpha^4 \,A^{sd}_{2;1,1}\,
e^{- \Omega_{1,2}\Lb Y;b\Rb}\,\,+\,\,2\,\beta^4\alpha^2
\,A^{sd}_{2;1,2}\,e^{- \h\Lb \Omega_{1,2}\Lb Y;b\Rb
+ \Omega_{2,2}\Lb Y; b \Rb\Rb}\,\,+\,\,\beta^6\,\,
A^{sd}_{2; 2,2}\,e^{- \Omega_{2,2}\Lb Y;b\Rb} \right\}. \nonumber
\eea
The cross section of the double diffractive
production is also a sum of the GW contribution, 
\beq \label{FSDD3}
\sigma^{GW}_{dd}\,\,=\,\,\int \,d^2 b \,\, \alpha^2\,\beta^2
\Big{|} A_{1,1}\, -\,2\,A_{1,2}\,+\, A_{2,2} \Big{|}^2,
\eeq
to which we add the term which is 
determined by the non GW contribution, 
\beq \label{FSDD4}
\sigma^{\mbox{nGW}}_{dd}\,\,=\,\,\int\,d^2 b\,
\left\{\alpha^4\,A^{dd}_{1,1}\,e^{- \Omega_{1,1}
\Lb Y;b\Rb}\,+ \,2 \alpha^2\,\beta^2 A^{dd}_{1,2}\,
e^{- \Omega_{1,2}\Lb Y;b\Rb}\,+\,\,\beta^4\,A^{dd}_{2,2}\,
e^{- \Omega_{2,2}\Lb Y;b\Rb}\,\right\}.
\eeq
  In our model  the GW sector can contribute to both low and high
diffracted mass, as we do not know the value of the typical mass for this 
mechanism, on the other hand, the non GW 
sector contributes only to
high mass diffraction.

\section{Results}

Using these formulae we fit the data with energies 
$W\,=\,\sqrt{s}\,>\,500\,GeV$, 
including the LHC data.
Our best set of parameters is presented in Table 1.  
In the same table we also show the values of the parameters that we 
obtained in our previous pre LHC fit,  which  included the data for 
($W \leq 1800\,GeV$).
 Although, the two sets of parameters differ somewhat, 
 the two key parameters, 
the Pomeron intercept $\Delta_\pom$ 
and the slope of the Pomeron trajectory $\alpha'_\pom$, 
remain almost the same.

\TABLE[ht]{
\begin{tabular}{|l|l|l|l|l|l|l|}
\hline
$\Delta_\pom $ & $\beta$ &  $\alpha^{\prime}_{\pom} (GeV^{-2})$& $g_1\,(GeV^{-1})$ &  $g_2\,(GeV^{-)}$ & $m_1$ \,( GeV)&
$m_2$\,(GeV)  \\ \hline
0.21 (0.2)& 0.46 ( 0.388) & 0.028 (0.02) & 1.89 (2.53)  &
61.99 (88.87)  &  5.045 (2.648 ) & 1.71 ( 1.37)
\\ \hline
$ \Delta_{\reg} $ & $\gamma $ &  $\alpha^{\prime}_{\reg}\,(GeV^{-2})$& $g^{\reg}_1\,(GeV^{-1})$ &
$g^{\reg}_2\,(GeV^{-1})$ & $ R^2_{0,1}\,(GeV^{-1})$& $G_{3\pom}\,(GeV^{-1})$ \\ \hline
-\,0.47 (-0.466) & 0.0045 ( 0.0033) & 0.4 (0.4) &  13.5 (14.5) & 800 (1343)  &
4.0 (4.0) & 0.03 (0.0173) \\ \hline
\end{tabular}
\caption{Fitted parameters for our model. In parenthesis we include the values that we obtained in our 
model without LHC data (see Ref.\cite{GLM}.
The quality of the fit is $\chi^2/d.o.f.$ = 0.86 (see the detailed explanation in the text.)
}
\label{t1}}

The comparison with the LHC data at $W$ = 7 TeV is shown in Table 2. 
Even though we are definitely below the data points for total and elastic 
cross 
sections\cite{TOTEM}, the corresponding large experimental error bars do 
not enable a definitive 
assessment of our predictions. Our predictions for $B_{el}$ and 
$\sigma_{inel}$ are compatible with the TOTEM data.   
TOTEM's $\sigma_{tot}$    
depends on the value of $\rho = \mbox{Re A}/\mbox{Im A}$, 
where $\mbox{Re A}$ and $\mbox{Im A}$ are the real and imaginary parts 
of the scattering amplitude $A$, 
and on $d\sigma_{el}/dt(t=0)$.

 In our model we assume that  Pomeron exchange  leads to a pure
 imaginary amplitude. Since, we expect that the real part of the amplitude
 will be much smaller than the  imaginary one, we can calculate the real
 part using a
 perturbative approach.

First, we notice that for one Pomeron exchange

\beq \label{REPOM}
\mbox{Re} A_\pom (s ,b )\,\,=\,\,\tan \frac{\pi \Delta_{\pom}}{2}\,\mbox{Im} A_\pom
\eeq
Having \eq{REPOM} in mind we can calculate the real part of the scattering 
 amplitude as follows
\beq  \label{RE}
\mbox{Re} A_{i k} (s ,b )\,\,=\,\,\mbox{Re}  \Omega_{i k}/2\,\exp\Big(
 - \frac{\Omega_{i k}\Lb s, b \Rb}{2}\Big)
\eeq
We use \eq{REPOM} to calculate $\mbox{Re}  \Omega_{i k}$.

Our value of $\rho $ is 
smaller than the COMPETE \cite{CUD} value 
which was used by TOTEM. The COMPETE $\rho$ fit  
is based on an extrapolation from data in the ISR-Tevatron range. 
The effect of our value of $\rho$ being smaller than the COMPETE value, 
implies a
 change in $\sigma_{tot}$  of less than 1\%.

In Table 3 we present our results for 
$\sigma_{tot}, \sigma_{el}, \sigma_{sd}, \sigma_{dd}, B_{el} ,B_{sd} $ and 
$\sigma_{inel}$.  
In parenthesis we put the values of these observables 
obtained from our previous fit. 
 The new fit gives higher values for the 
single and double diffraction, while changes in all other 
observables are rather small. It is interesting to note that 
Block and Halzen \cite{BLHA} have ``converted"\cite{BLHA} 
a recent measurement of
the Pierre Auger Observatory collaboration of proton-air collisions with
$\sigma^{p-air}_{in}$ at W=57 $\pm$ 6 TeV to $\sigma_{in}$ 
for proton-proton collisions, and obtain the value of 
$\sigma^{pp}_{in} = 90 \pm 7(stat) \pm 1.5 (Glauber( +9/-11(syst)\,$mb
and predict $\sigma^{pp}_{tot} = 134.8 \pm 1.5 \,$ mb at this energy.

It is interesting to note that at W = 57 TeV, we have that the ratio 
$\frac{\sigma_{in}}{\sigma_{tot}}$ = 0.74, while Block and 
Halzen\cite{BLHA}  have the
 value 0.69, both far from 
the black disc value of 0.5. 

 In Table 3 we show that we obtain the same value 
for the inelastic cross section, while for the total cross 
section we have 122 $\,$ mb which is smaller than has been 
advocated in Ref.\cite{BLHA}.

In \fig{fit} we show  our present fit in the  energy range
$ 20 \leq W \leq  1800 GeV$  energy range. We see that the description is
 not as 
good 
as in our previous paper\cite{GLM}, nevertheless, our model 
 describes the main features of low energy data as well.

{\small
\TABLE{
\begin{tabular}{|l |l|l|l|l|}
\hline \hline
W & $\sigma^{model}_{tot}$ &  $\sigma^{exp}_{tot}$ &  $\sigma^{model}_{el}$&
 $\sigma^{exp}_{el}$\\
\hline
7 TeV~~~ & 94.2 mb~~~~& TOTEM: 98.3 $\pm$0.2$^{st}$ 
$\pm$2.8$^{syst}$mb~~~~~~ & 22.9 mb~~~~~ & TOTEM: 24.8$\pm 0.2^{st} \pm 
1.2^{syst}$mb~~~~~~\\
\hline
\end{tabular}
\begin{tabular}{|l|l|l|l|l|}
\hline
 W & $\sigma^{model}_{in}$& $\sigma^{exp}_{in}$&$B^{model}_{el}$& $B^{exp}_{el}$\\
\hline
7 TeV & 71.7 mb & CMS: 68.0$\pm2^{syst}\pm 2.2^{lumi}\pm 4^{extrap}$ mb & 19.8 $GeV^{-2}$&
TOTEM: 20.1$\pm 0.2^{st} \pm 0.3^{syst}\,GeV^{-2}$\\
 &  &  ATLAS: 69.4$\pm 2.4^{exp}  \pm 6.9^{extrap}$ mb & &\\
& &  ALICE: 72.7 $\pm 1.1^{model}  \pm 5.1^{extrap}$ mb & & \\
& & TOTEM: 73.5  $\pm 0.6^{st}  \pm 1.8^{syst}$ mb& & \\
\hline
\end{tabular}
\begin{tabular}{|l |l|l|l|l|}
\hline
W & $\sigma^{model}_{sd}$ &  $\sigma^{exp}_{sd}$ &  $\sigma^{model}_{dd}$& $\sigma^{exp}_{dd}$\\
\hline
7 TeV & 10.5${}^{GW}$ +  2.6${}^{nGW}$ mb~ & ALICE : 14.16  $\pm$ 3 mb  & 5.98${}^{GW}$  + 1.166${}^{nGW}$ mb ~~~~~~~& ALICE: 8.86 $\pm$ 3 mb~\ \\
\hline\hline
\end{tabular}
\caption{Comparison of the predictions of our model with the experimental data
 at W= 7 TeV.}
\label{t2}

}
}
~

~
\TABLE{\small
\begin{tabular}{|l|l|l|l|l|l|}
\hline
$\sqrt{s}$  \,\, TeV
& 1.8
&  7
&  14
& 57 \\
\hline
 $\sigma_{tot}$ mb

&  75.6 (74.4)
& 94.2 (91.3)
& 104.0 (101)
&  122.0  \\
\hline
$\sigma_{el}$ mb
& 18.2 (12.5)
&  22.9 (23)
&  26.1 (26.1)
& 31.1   \\
\hline
 $\sigma_{sd}(M \leq M_0)$ mb
 &
& 10.5 + $(2.6)^{nGW}$ (10.2)
&  11.2 + $(3.32)^{nGW}$(10.8)
&  12.8 + $(3.91)^{nGW}$   \\
\hline
$ \sigma_{sd}( M^2 < 0.05s)mb $&
8.97+ $(1.95)^{nGW}$ (8.87)
&10.5 + $(3.94)^{nGW}$ (10.2)
&  11.2 + $(5.58)^{nGW}$(10.8)
&  12.8 + $(8.19)^{nGW}$   \\
\hline

$\sigma_{dd}$ mb
& 5.56 + $(0.369)^{nGW}$ (4.46)
& 5.98 + $(1.166)^{nGW}$ (6.46)
& 6.55  + $(1.5)^{nGW}$ (6.65)
& 8.61 + $(4.9)^{nGW}$   \\
\hline
$B_{el}\;GeV^{-2}$
&  17.6 (16.1)
& 19.8 (19.3)
& 21.2 (20.5)
& 23.8   \\
\hline
$B^{GW}_{sd}\;GeV^{-2}$
&  6.36
& 8.01
& 8.78
& 10.4  \\
\hline
$\sigma_{inel}$ mb
& 57.4
&  71.7
& 77.9
&  90.9
 \\
\hline \hline
\end{tabular}
\caption{ Predictions of our model for different energies $W$. $M_0$ is
 taken to be equal to $200GeV$ as ALICE measured the cross section of
 the diffraction production with this restriction.}
\label{t3}}

\FIGURE[h]{
\begin{tabular}{c c}
\epsfig{file=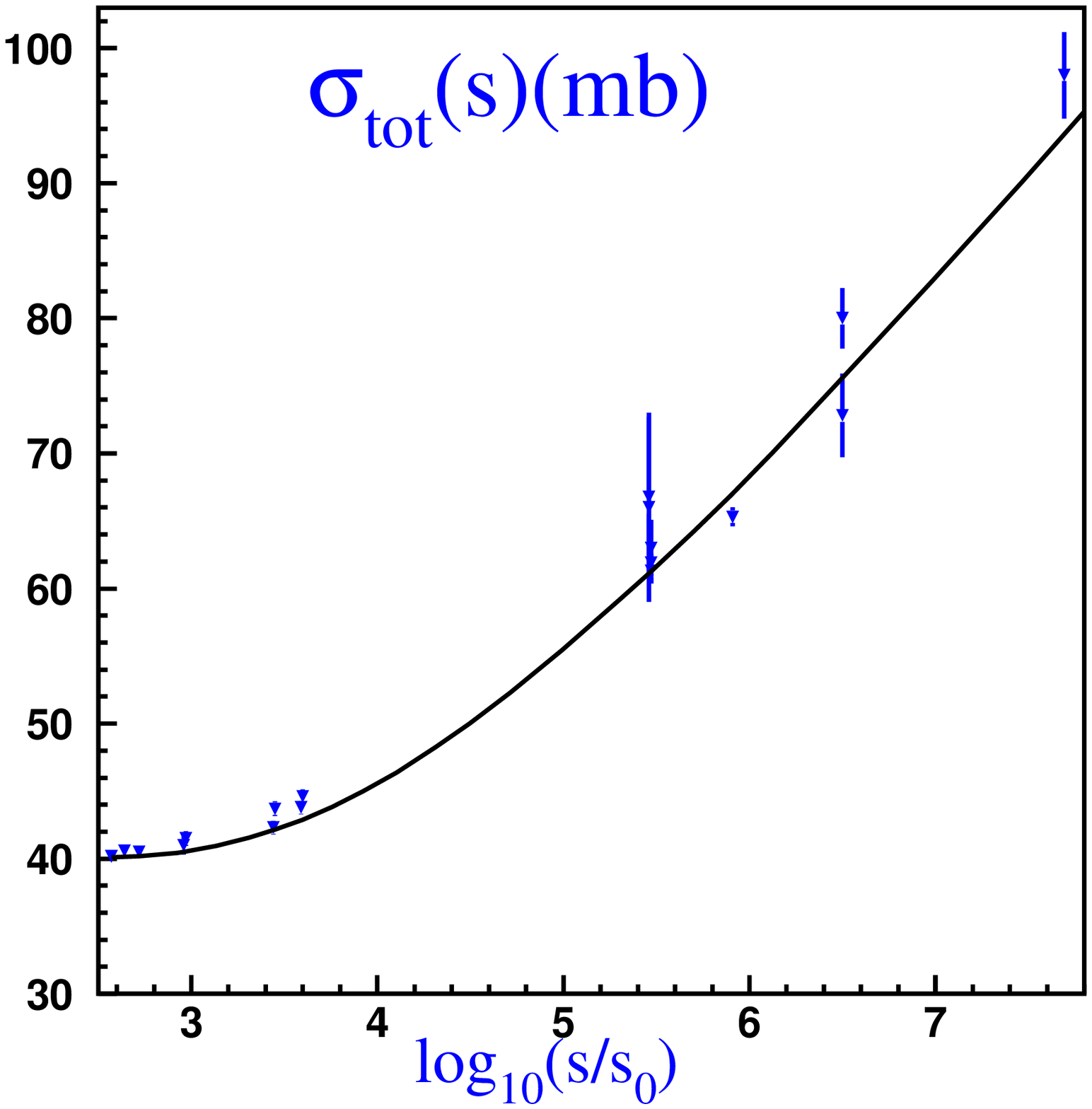,height=50mm,width=65mm}
 &\epsfig{file=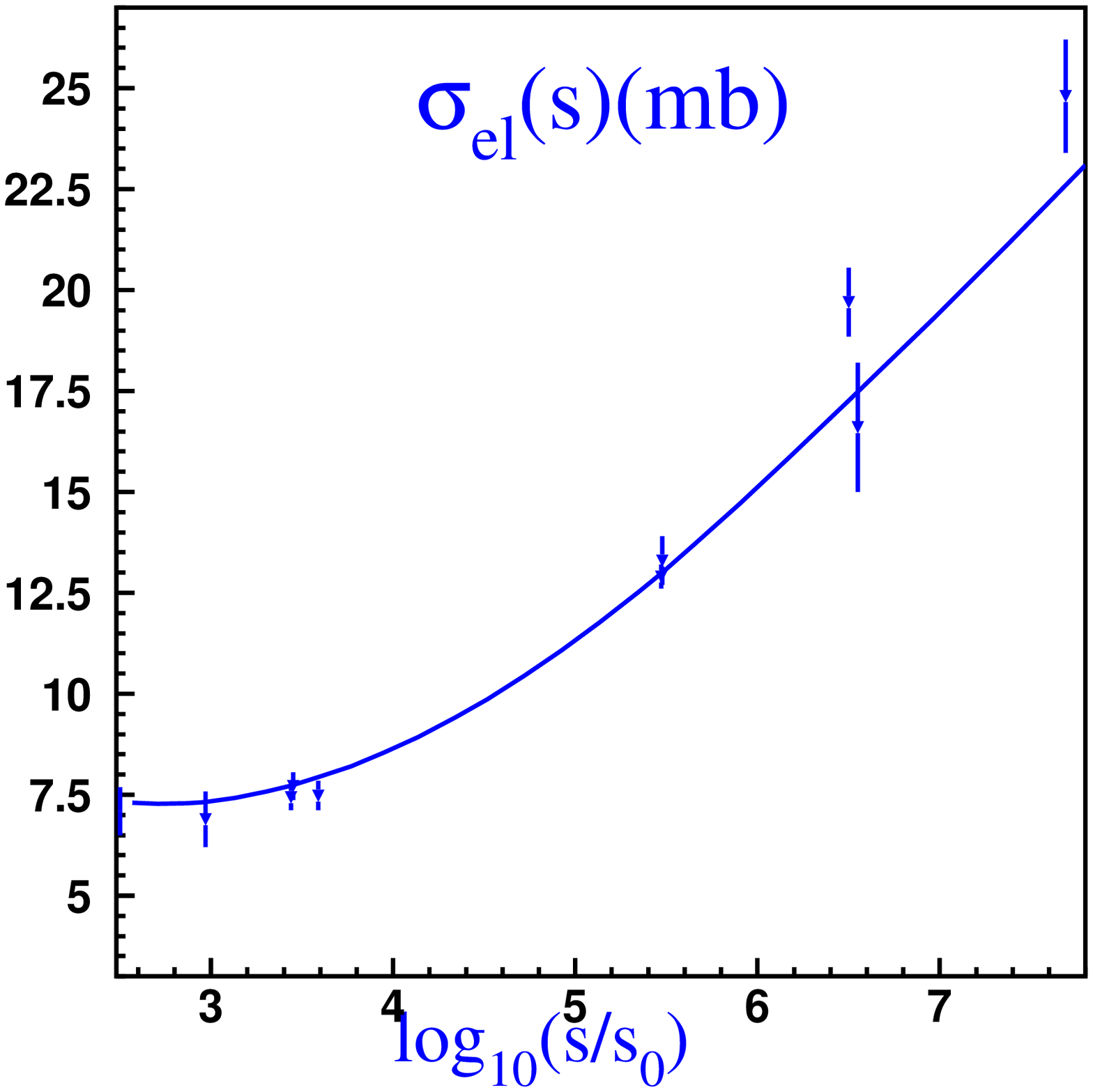,height=50mm,width=65mm}\\
\fig{fit}-a & \fig{fit}-b\\
\epsfig{file=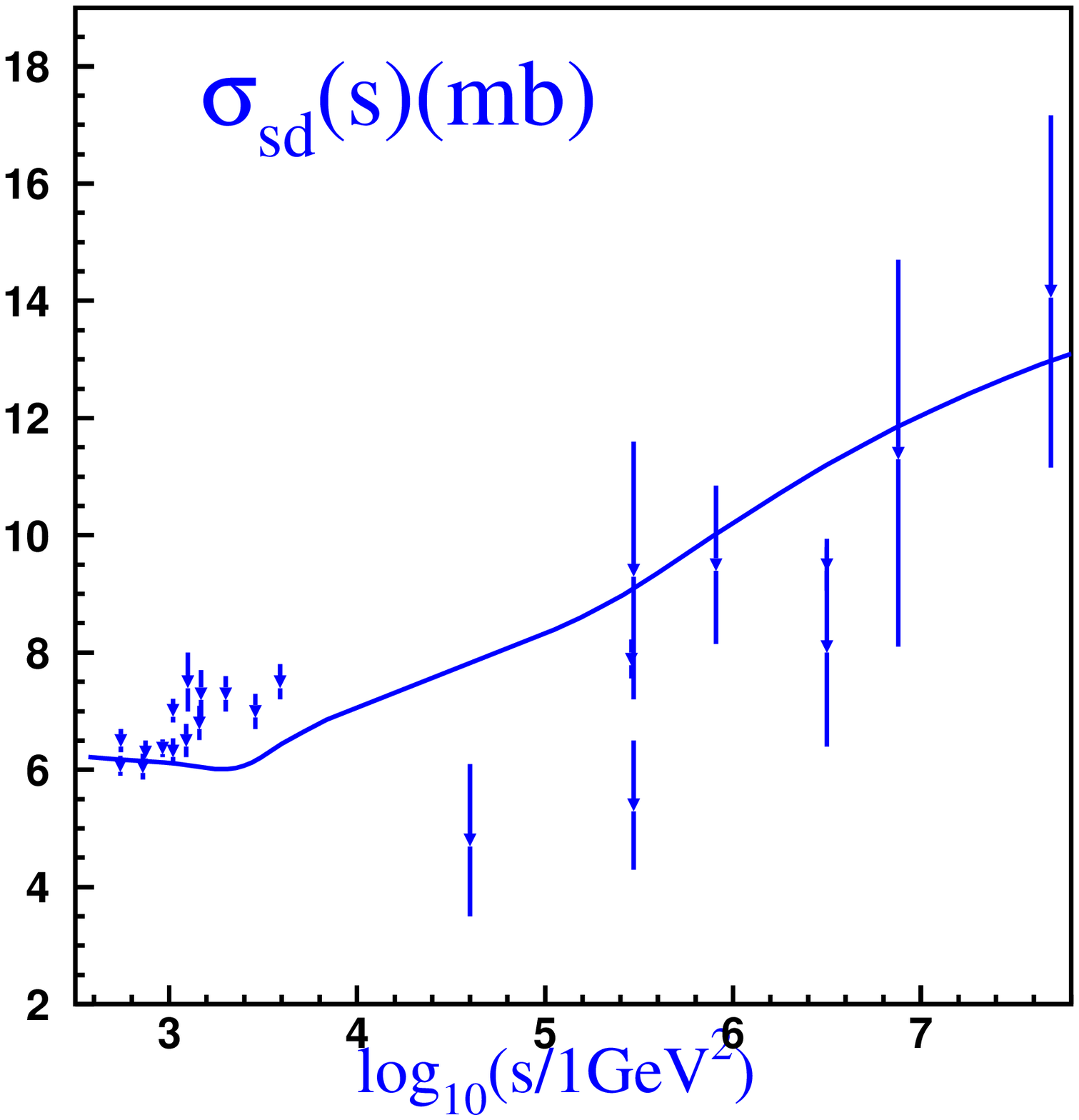,height=50mm,width=65mm}
 &\epsfig{file=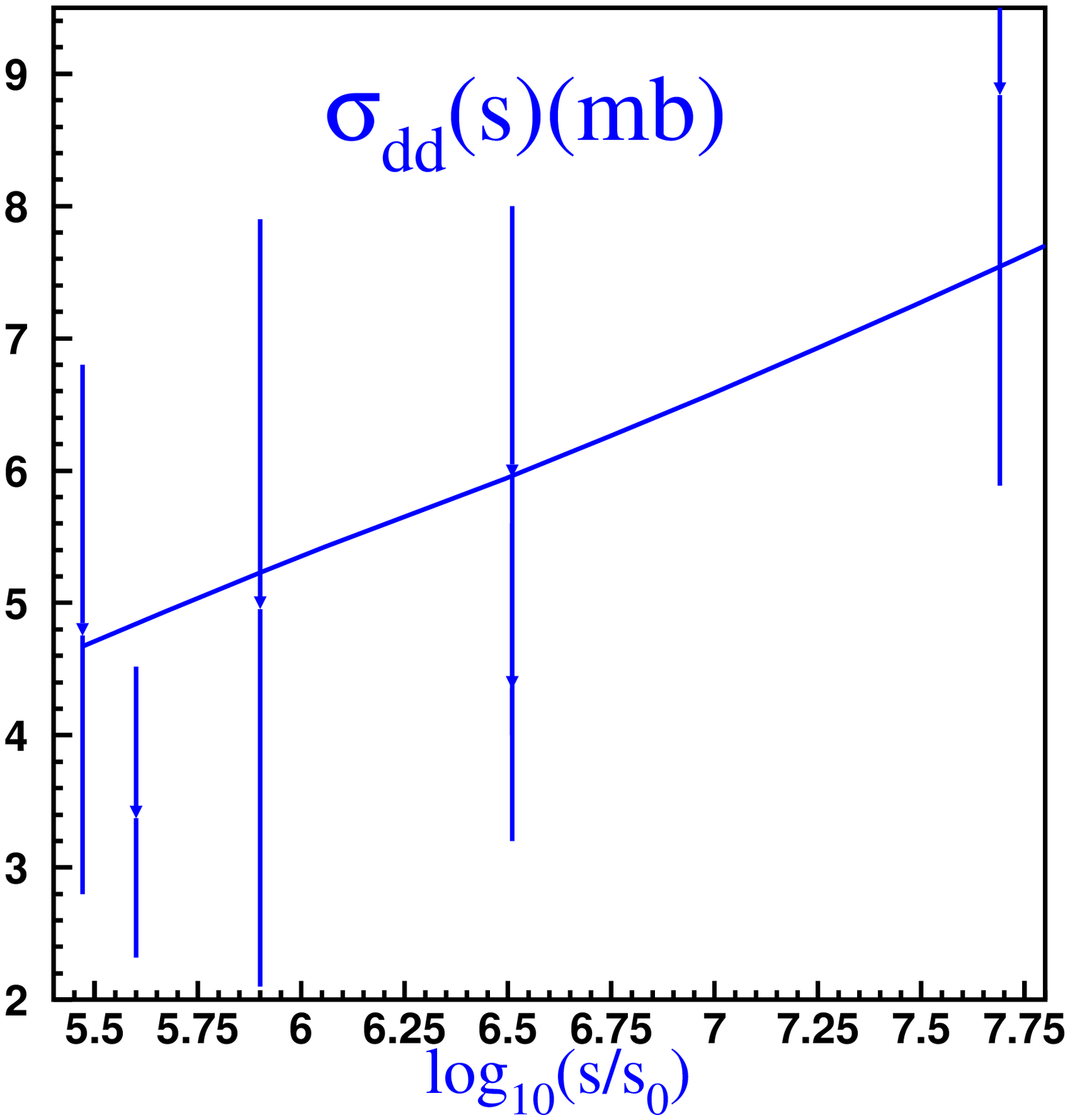,height=50mm,width=65mm}\\
\fig{fit}-c & \fig{fit}-d\\
\epsfig{file=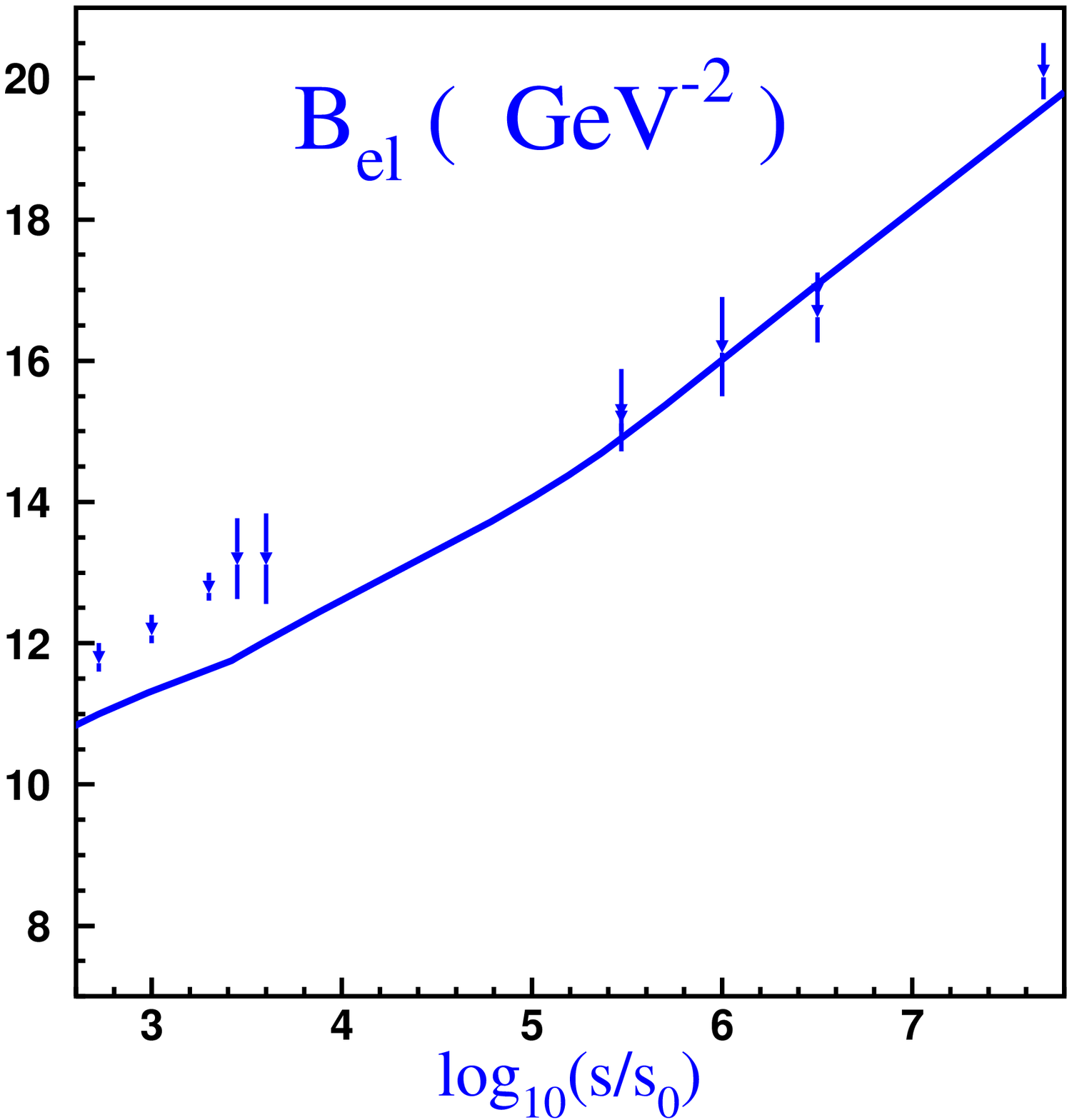,height=50mm,width=65mm}
 &\epsfig{file=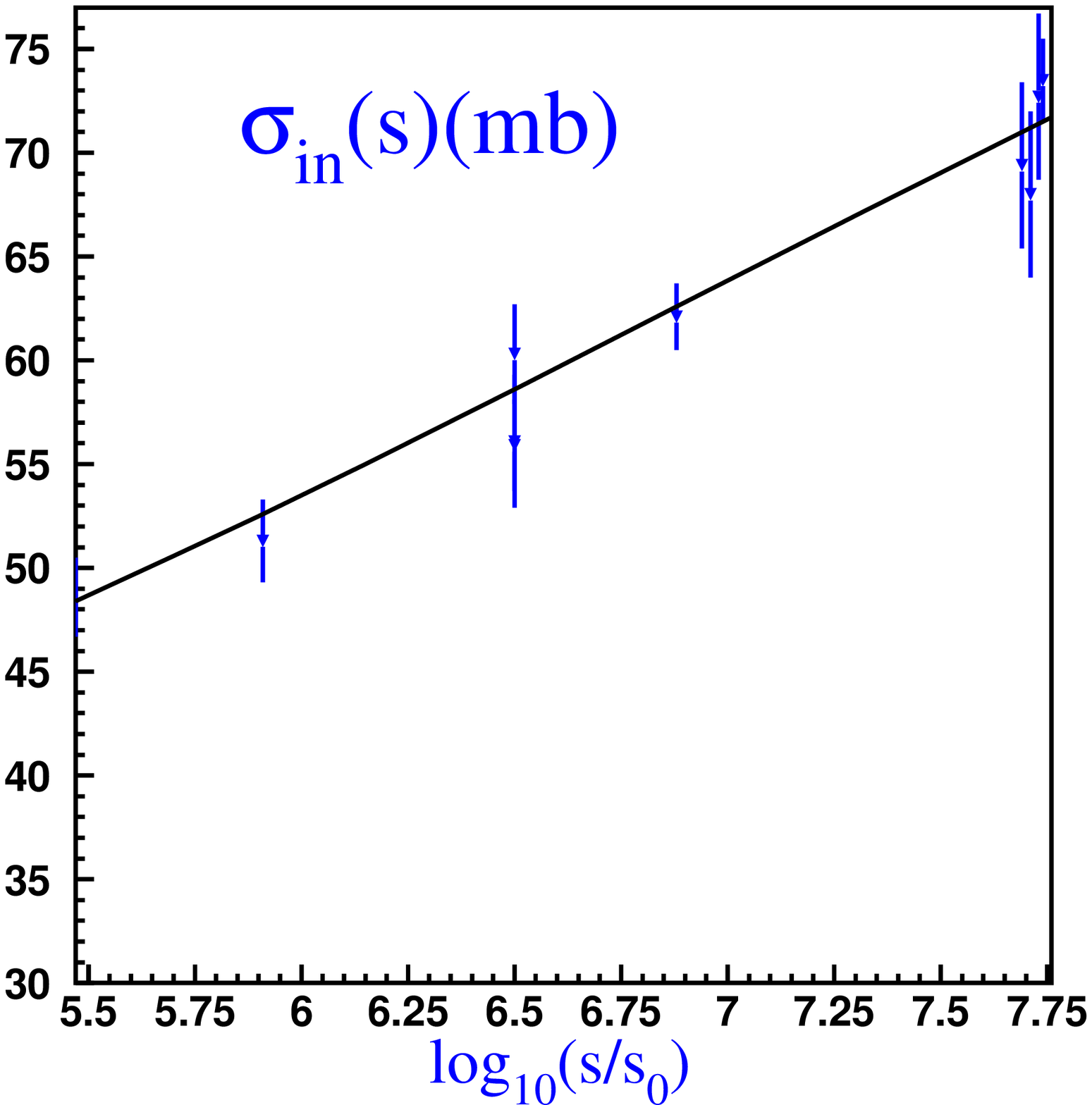,height=50mm,width=65mm}\\
\fig{fit}-e & \fig{fit}-f\\
\end{tabular}
\caption{Comparison with the experimental data the energy behaviour
 of the total (\protect\fig{fit}-a),
elastic  (\protect\fig{fit}-b), single diffraction  (\fig{fit}-c),
 double diffraction (\protect\fig{fit}-d) and inelastic (\protect\fig{fit}-f) cross sections
 and elastic slope( \protect\fig{fit}-e) .
 The solid lines show our present fit . The data has been taken from
 Ref.\protect\cite{PDG} 
for energies less than the LHC energy. At the LHC energy for total and
 elastic cross 
section we use TOTEM data\protect\cite{TOTEM} and for single and double
 diffraction 
cross sections are taken from Ref.\protect\cite{ALICE}.
}
\label{fit}}

 In \fig{am} we plot the amplitudes $A_{i k}$ at different energies as
 a function of the impact parameter $b$.  The structure of the b and s
 dependence, remains essentially the same as in our 
previous descriptions.
 However, the amplitude $A_{11}$ turns out to be smaller than that 
obtained previously  \cite{GLM}.  This amplitude does not reach the unitarity
 bound even at W = 57 TeV.
Recall that the unitarity black disc bound at  a
 given $(s, b)$ is only reached  when $A_{11}(s, b) = A_{12}(s,b) =
A_{22}(s,b) = 1$. As seen in \fig{am} this condition has not yet been 
satisfied   at  W = 57 TeV.
 In some sense, our model
gives an example of why it is dangerous  to apply the simple formulae 
that are
 valid for the black disc 
 regime, for extracting  information on the high energy scattering from 
the experimental data. In particular, as we
 have discussed 
 at W = 57 TeV,  we obtained  $\sigma_{tot} = 122mb$ while $\sigma_{el} = 
31 mb$ 
 in clear violation of the relation $\sigma_{el} = \sigma_{tot}/2$
  for a black disc.

The impact parameter dependence shown in \fig{am} reflects 
 in the  $t$ dependence of the elastic cross section shown in \fig{dsdt}.
 One can see that we are in agreement with the experimental results for 
both 
 Tevatron and  LHC energies in the forward cone  for -t $\leq \, 0.5 \, 
GeV^{2}$. In \fig{dsdt} we show our  prediction for $W=14\,TeV$.

A significant check for our model will be the measurement of 
$t$-dependence for
 $d \sigma_{sd}/dt$. As shown in 
 Table 3 we predict quite a small slope $B_{sd}$  for the Good-Walker 
contribution
 of the single diffractive production.
For large mass diffraction due  to triple Pomeron interaction we expect
 that the slope will be about $B_{el}/2$, since the slope
 for triple Pomeron vertex is rather small $ \leq 1 \,GeV^{-2}$. For 
single
 diffraction we obtain a value of the slope 
  $ B_{sd} = 8.01 \,GeV^{-2}$, (at t =0), this is compatible with the 
preliminary result
 of TOTEM as presented by Risto Orava \cite{RISTO}.

As  is well known, the survival probability for the large rapidity gap
 crucially depends on the $b$-dependence of the amplitude. 
 To check our $b$-dependence as well as 
for completeness of our presentation, we calculate the survival 
probability 
$S^2$ for a large rapidity gap using the general formulae of 
Ref.\cite{GLMSP}. We obtain
\beq \label{SP}
S^2 = 9.76 \% ( 10\%)\,\,\mbox{at}\,\,W=1.8\, TeV;\,\,\,\,\, 
S^2 = 5.32 \% ( 6.3\%)\,\,\mbox{at}\,\,W= 7\, TeV; \,\,\,\,\,
S^2 =  3.66 \% ( 4.4\%)\,\,\mbox{at}\,\,W=14\, TeV;
\eeq
where we indicate in  parentheses the values of the survival 
probability calculated
in our previous paper \cite{GLMSP}. 
 Naively, we could expect that $S^2$ would be larger than in our previous 
 model since the larger transparency of $A_{11}$ in this approach 
 (see \fig{am}).  From \fig{am} we see that $A_{11}(b)$ in
 this model is smaller,  and decreases faster than this amplitude in
 our previous model. Both these effects lead to an increase of the
 survival probability due to the Good-Walker mechanism. However,
 the considerable increase of the values for $G_{3 \pom}$ and $\gamma$
 lead to stronger screening, due to non Good-Walker mechanism, mostly due
 to contribution of the enhanced diagrams (see  Refs.\cite{GLM,GLMSP}).
 The second effect that leads to the decrease of $S^2$, is the value of
 $\alpha$ in this model, which is smaller than in our previous approach.
 Recall that $ \alpha^2 = 1 - \beta^2$, and  $S^2 \propto \alpha^4$,
since the contribution of $A_{12}$ and $A_{22}$ to the value of the
 survival probability is negledgibly small due to large $\Omega_{12}$
 and $\Omega_{22}$ at  small $b$.

\FIGURE[h]{\begin{tabular}{c c }
\epsfig{file=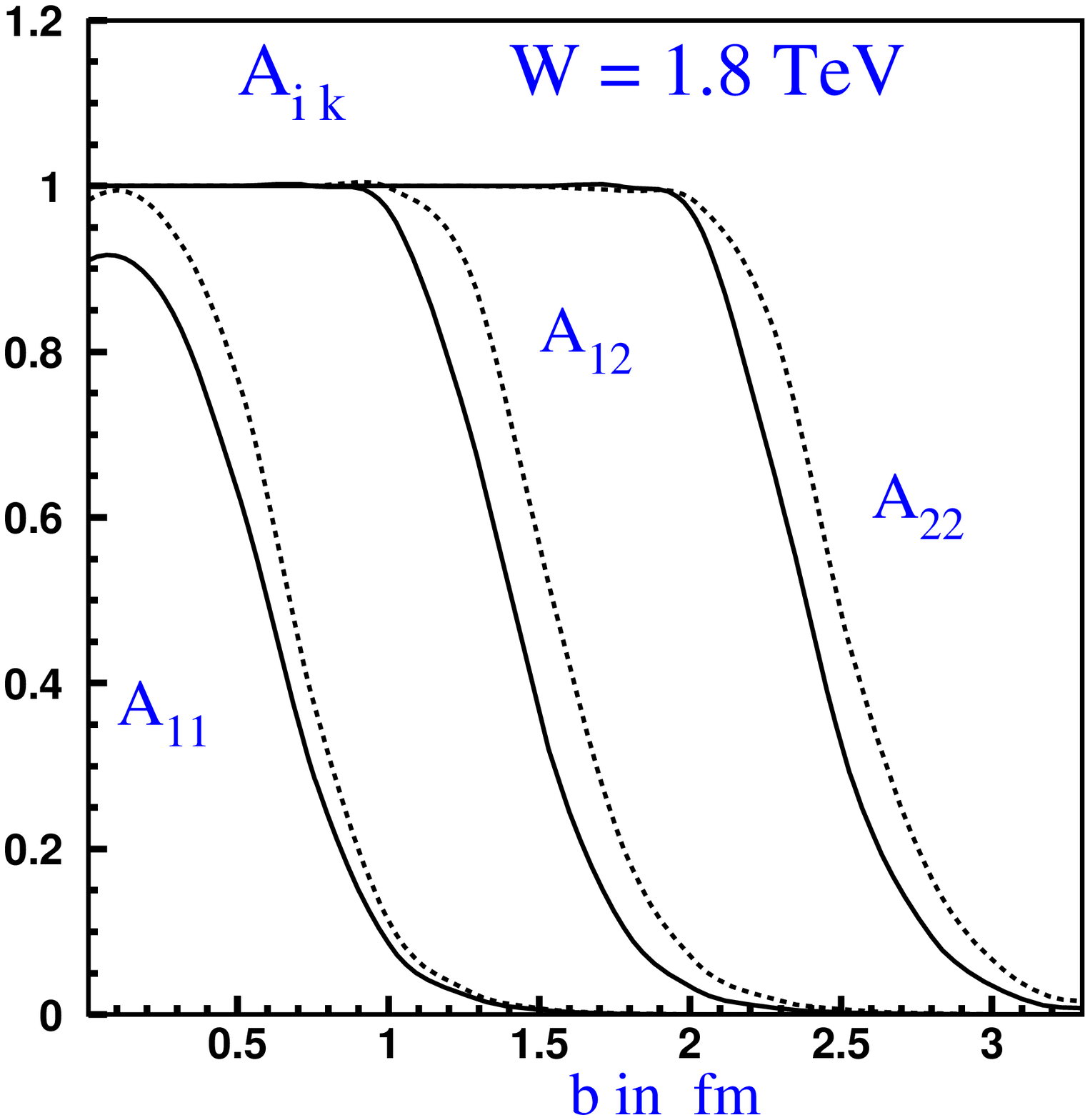,width=70mm}&\epsfig{file=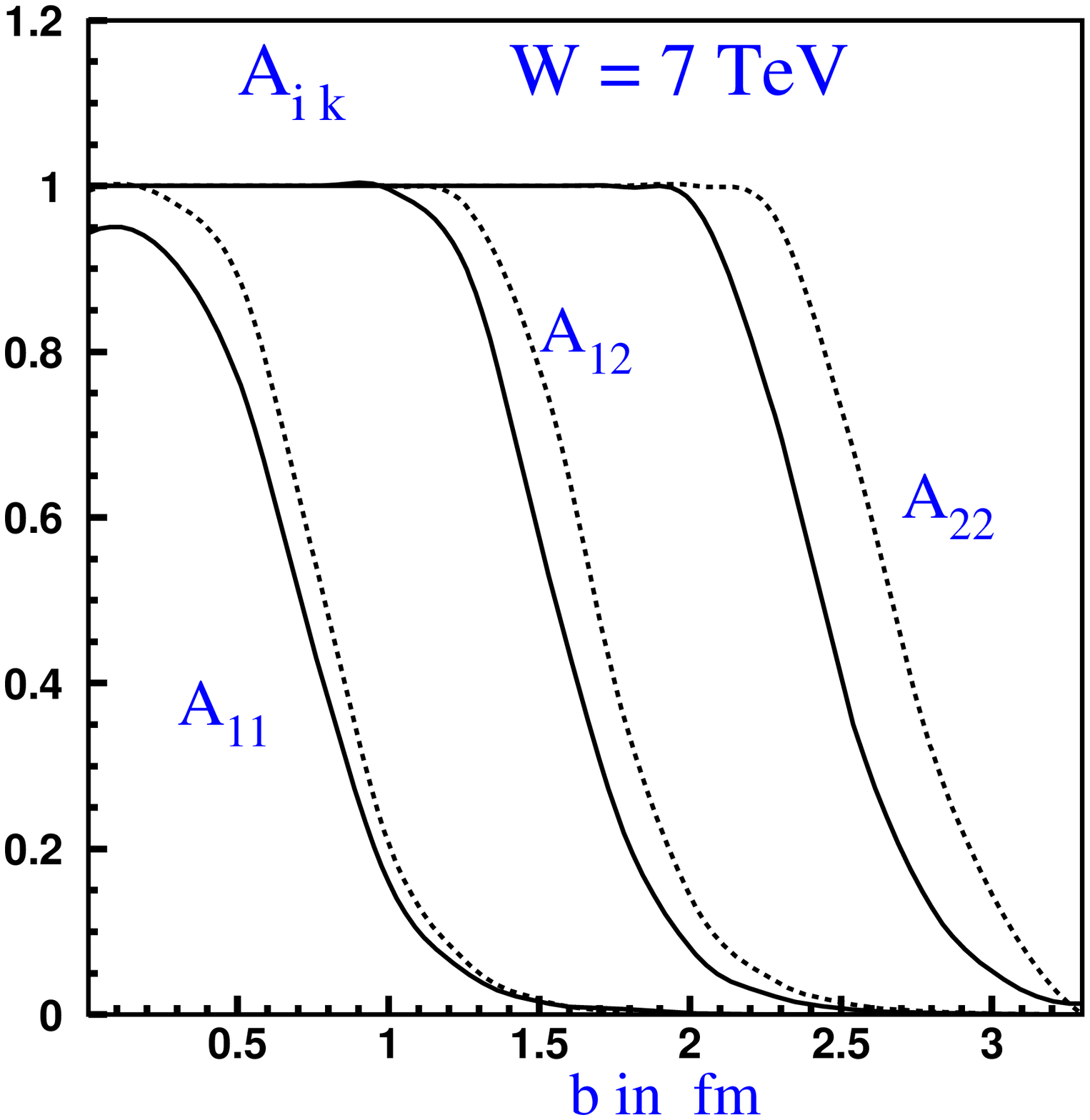,width=70mm}\\
\fig{am}-a &\fig{am}-b\\
 \epsfig{file=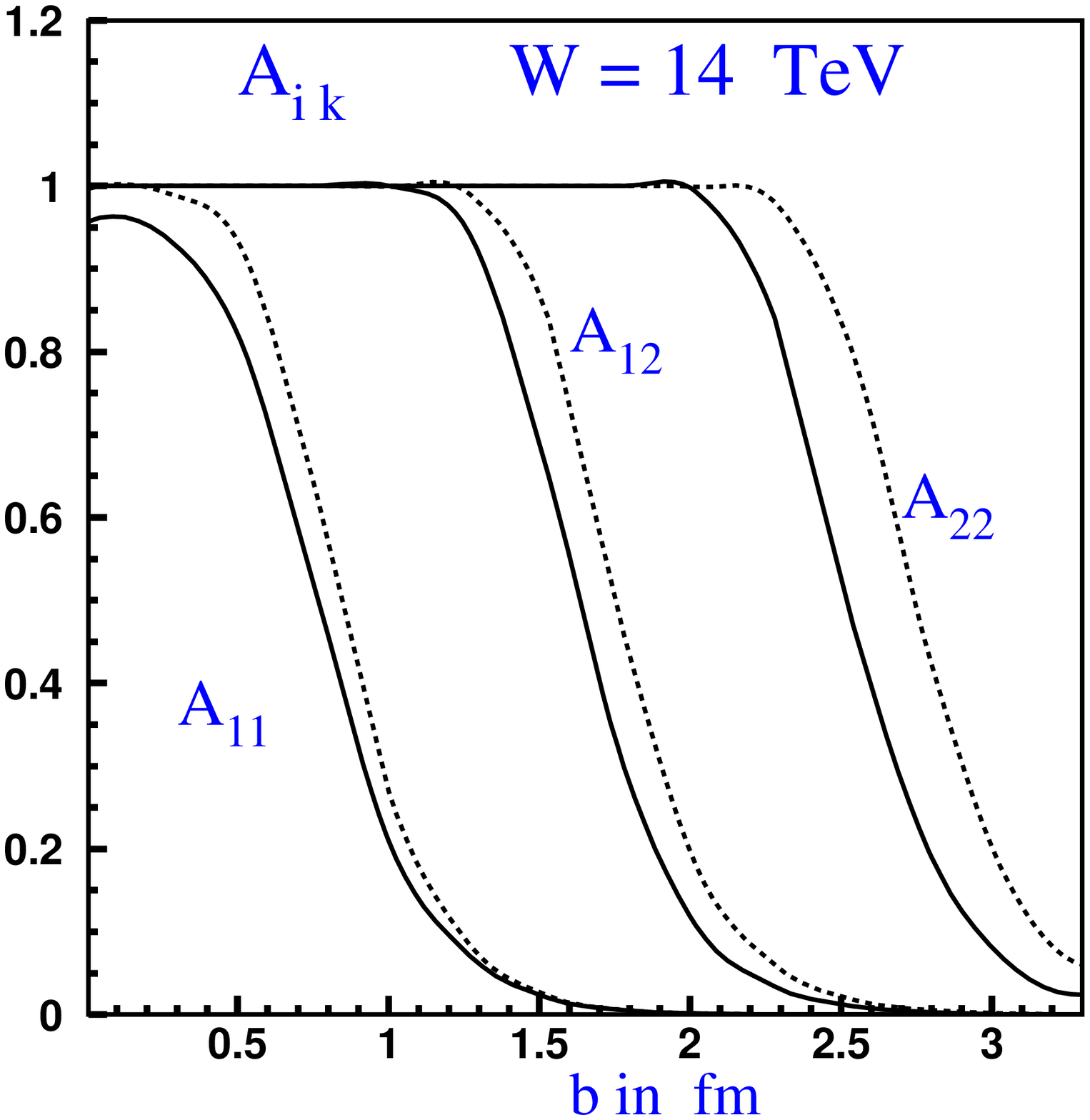,width=70mm}&\epsfig{file=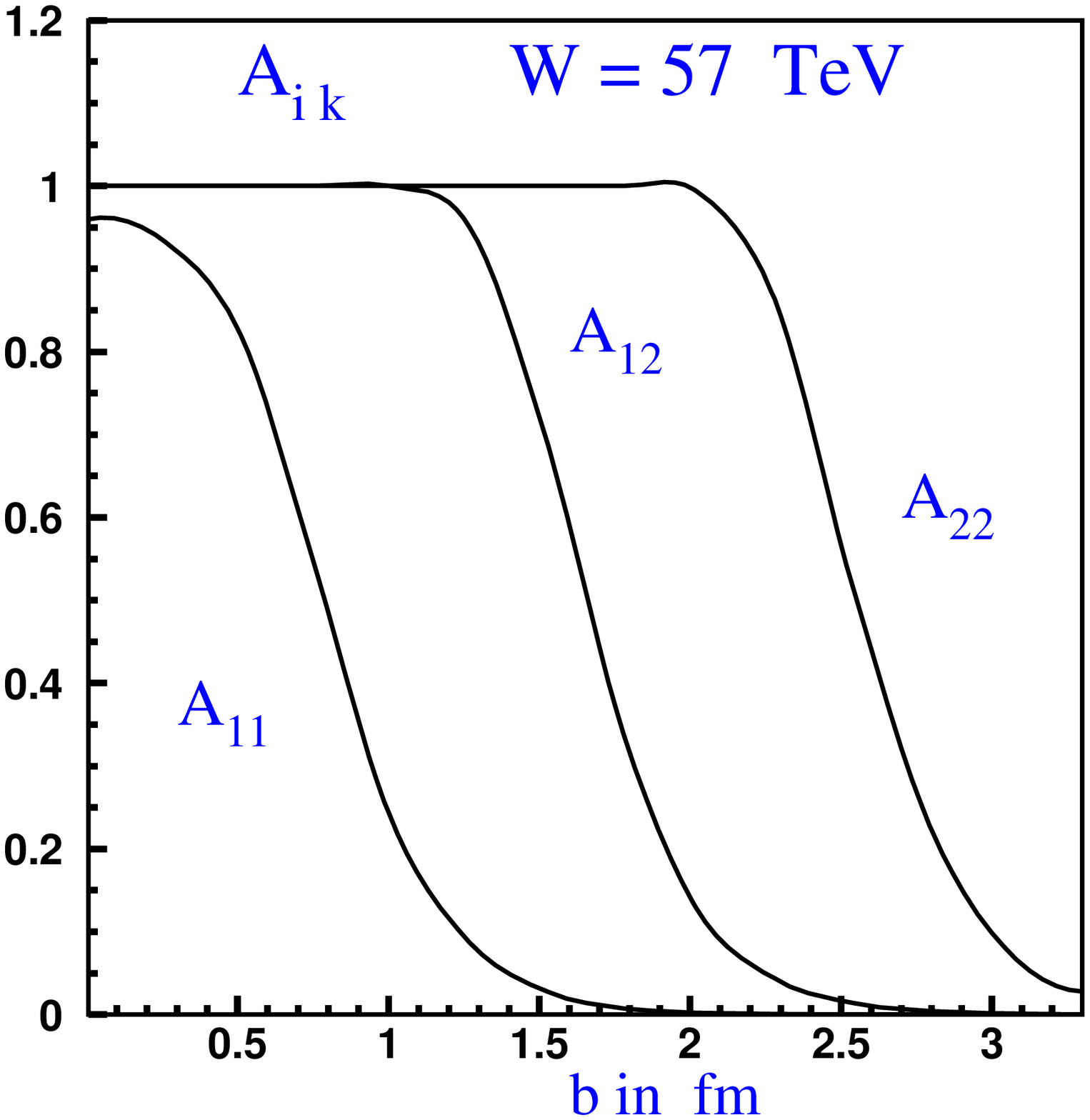,width=70mm}\\
\\
\fig{am}-c &\fig{am}-d\\
\end{tabular}
\caption{ The impact parameters dependence of amplitudes $A_{i k}$ at
 different energies. 
The solid lines are the results of this model, the dashed line are
 the amplitudes given by 
the model of Ref.\cite{ GLM}.}
\label{am}
}

\FIGURE[t]{
\centerline{\epsfig{file=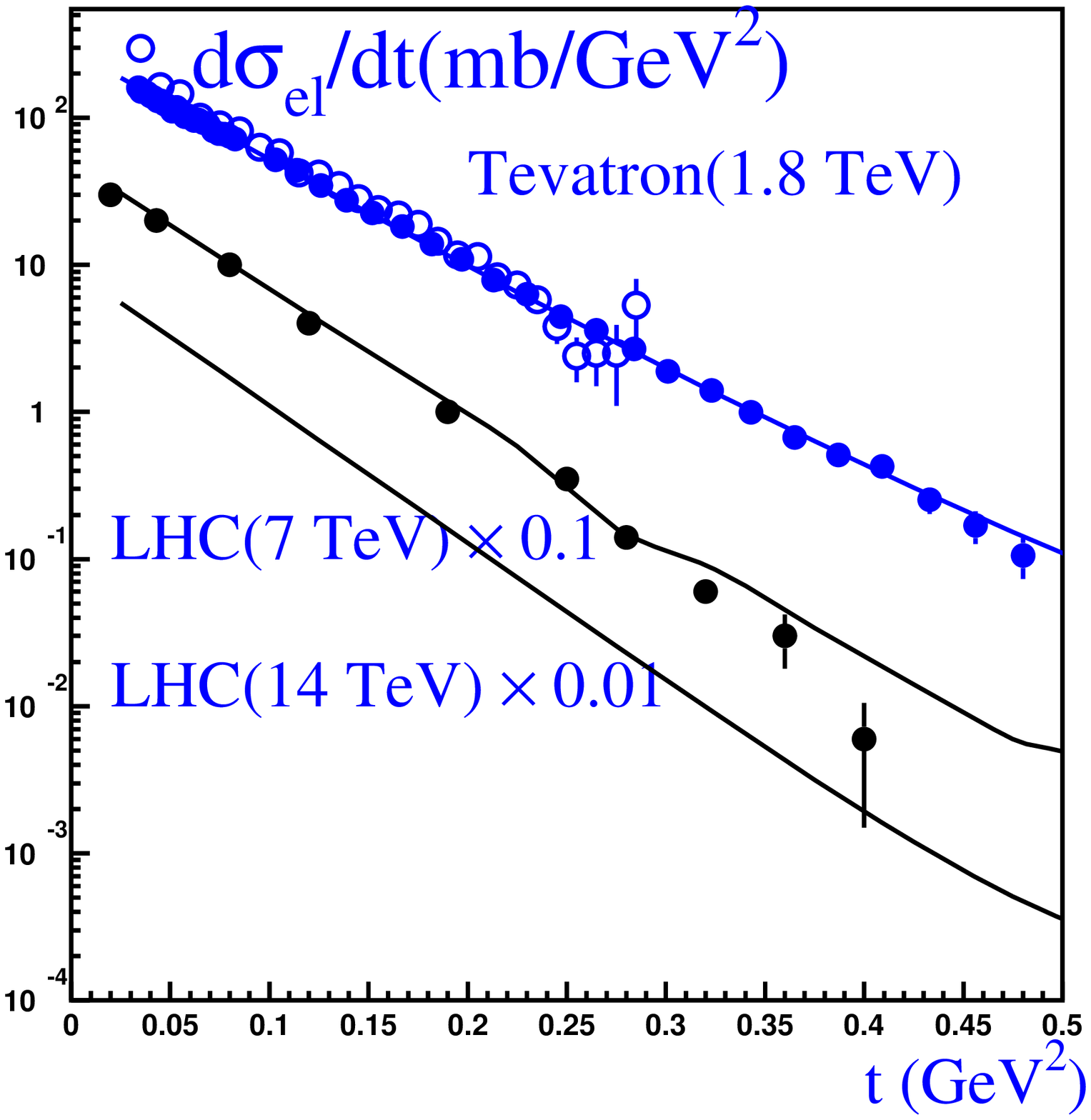,width=100mm,height=80mm}}
\caption{ $d \sigma_{el}/dt$     versus $|t|$     at Tevatron (blue curve and data)) and LHC ( black curve and data) energies ($W = 1.8 \,TeV $ and $7 \,TeV$ respectively) The solid line without data  shows our prediction for $W=14\,TeV$.
 Data from Refs.\cite{E710,TOTEM}.}
\label{dsdt}}

\section{Conclusions}

We are able to reproduce the LHC experimental 
data within the present experimental errors and uncertainties.
Therefore, we found a positive answer to the question formulated 
 in the introduction: our model without any changes except the new set 
 of parameters is able to describe the LHC data with a good accuracy. 
It is instructive to note, that we did not spoil the reasonable 
description of the low energy data ($W \leq 500\,GeV$) being the only model 
 on the market that is able to describe all available data on high energy 
 scattering.  We would like to stress
that our predictions for $\sigma_{tot}$ and $\sigma_{el}$ are
below those published by TOTEM \cite{TOTEM}, while our values for
$\sigma_{inel}$ and $B_{el}$ are in agreement with the published
experimental
values.  If the published errors will be reduced while
the central values will remains approximately the same, our approach
will need essential improvements and new ideas.
  
As one can see from Table 1 the changes in parameters are not dramatic but 
 three parameters: $m_1,\gamma$ and $G_{3\pom}$, turns out to be 
  two to three  times larger that in our previous approach \cite{GLM}.
 These changes are driven by the LHC data and they  deserve a discussion.
  The increase in $\gamma$ and $G_{3\pom}$ is a direct consequence
 of the large diffractive cross section measured at the LHC.  At first sight
 we could increase them even more to obtain a better description of the
 of the high mass diffraction production . Unfortunately, this is not true 
as
 the sum of enhanced and net diagrams  depend strongly  on the value 
of
  $\gamma$ and $G_{3\pom}$, leading
to a resulting decrease of the diffractive cross sections at larger values of
 these parameters.

The most striking  manifestation of larger value of $m_1$ we see in
 \fig{am} which shows that at ultra high energy the amplitude
 $A_{11}\Lb b = 0 \Rb$ is still less that unity. The slow increase to
 the unitarity limit in $A_{11}$ amplitude is a typical feature of our approach.
 However, the fact that this amplitude is less than 1 for 
 $b=0$ even at $W=57 \,GeV$ is certainly the consequences of the 
 LHC data in the framework of our model.

There are two lessons that we can learn from this fit. The first is that 
  if the
 TOTEM collaboration will confirm their results for $\sigma_{tot}$ and  
 $\sigma_{el}$, we will need to
 reconsider the main physical ideas on which our approach is built.
 The second one, is the fact that one has to be very careful
when considering the concept
 of the black disc regime.

\section{Acknowledgements}

This research was supported by the Fondecyt (Chile) grant 1100648. 

\end{document}